# The Secure Generation of RSA Moduli Using Poor RNG

Dr. George Purdy, University of Cincinnati

We discuss a procedure, which should be called Lenstra's fix, for producing secure RSA moduli even when the random number generation is very poor.

RSA is uniquely vulnerable to low entropy random number generation. If n = pq and n' = pq' are two (public) moduli, then the computation gcd(n,n') = p factors both moduli and totally compromises the security of both systems.

Following a suggestion of A. K. Lenstra and his coauthors in [1] we present an algorithm for generating p and q that avoids this attack without changing the method of random number generation.

If the probability is P that in the world two random primes p and p' are generated the same, then the probability that n = n' will only be $P^2$. It is much more likely that p = p' and q ≠ q', in which case gcd(n,n') = p factors n and n'

The damage done when n = n' is incomparably less. The owners of n and n' can access each other's accounts, but they are safe from attacks by others. Moreover, this is the same risk that any other cryptosystems face.

The proposal is to generate p randomly and then choose q = f(p,k), where f(p,k) is the function

$f(p,k) = 1 + [2^{2k}/p]$. Here [x] denotes the integer part of x. For example, [3.14] = 3.

**Algorithm 1**

1. Generate random prime p, $1+2^k < p < 2^{k+1}$.
2. Compute $q = f(p) = f(p,k)$.
3. Test q for primality.
4. If q fails to be prime, then goto 1.
5. Put $n = pq$.

The algorithm takes average time $O(k^2)$ instead of the usual $O(k)$.

We require the following properties of $f(p,k)$:

**Theorem 1**
1. If $1+2^k < p < 2^{k+1}$, then (1a) $f(p,k) < 2^k$ and (1b) $f(p,k) > 2^{k-1}$.
2. The function $f(p) = f(p,k)$ is bijective in the range $1+2^k < p < 2^{k+1}$. In other words, if $f(p) = f(p')$, then $p = p'$.

**Theorem 2**
If (1) and (2) are true, and if $n = pq$, $n'=p'q'$ and $\gcd(n,n') > 1$, then $n = n'$.

**Proof**
Suppose that $\gcd(n,n') > 1$.
Case 1. If $p = p'$, then $q = f(p) = f(p') = q'$, and therefore $n = n'$.
Case 2. If $q = q'$, then by (2) $p = p'$ and therefore $n = n'$.
Case 3. If $p = q'$ or $p' = q$. This cannot happen because by (1) all of the q's are less than $2^k$ and all of the p's are greater than $2^k$ by assumption. So no q can be equal to any p.

**Proof of theorem 1**
We first prove (1a), that $f(p,k) < 2^k$.

Case 1.
If $2^{2k}/p < 2^k - 1$, then $f(p) = 1 + [2^{2k}/p] < 1 + 2^{2k}/p < 2^k$, which is (1a).

Case 2. If $2^{2k}/p \geq 2^k - 1$. Then $2^{2k}/p > 2^k - 1$, since $2^{2k}/p$ is not an integer. Multiplying by p we obtain $p(2^k - 1) < 2^{2k}$. Hence $p(2^k - 1) \leq 2^{2k} - 1$. Hence $p \leq (2^{2k} - 1)/(2^k - 1) = 2^k + 1$. But by the hypothesis $p > 2^k + 1$, so case 2 cannot occur.

Proof of (1b), that $f(p) > 2^{k-1}$.
Now $f(p) = 1+[2^{2k}/p] \geq 2^{2k}/p$, and by the hypothesis $p < 2^{k+1}$. Hence $f(p) > 2^{2k}/2^{k+1} = 2^{k-1}$.

**Proof of (2)**
Suppose $f(p,k) = f(q,k)$. Then $[2^{2k}/p] = [2^{2k}/q] = r$, say. Hence $r = 2^{2k}/p - \varepsilon = 2^{2k}/q - \varepsilon'$, where $0 < \varepsilon, \varepsilon' < 1$.
Hence $|2^{2k}/q - 2^{2k}/p| = |\varepsilon - \varepsilon'| < 1$. Multiplying by pq we get $|2^{2k}p - 2^{2k}q| < pq$. But by the hypothesis $p < 2^{k+1}$, and by (1a) $q < 2^k$. Hence $|2^{2k}p - 2^{2k}q| < pq < 2^{2k+1}$, and $|p-q| < 2$. But p and q are odd. Hence $p = q$.

**Remarks**
The tacit assumption in all of this is of course that *everyone* generating keys will use this system.

Let's consider what happens if the routers made by Company A use the above algorithm, but the routers of Company B just choose p and q independently.

If company A issues $n = pq$, then $q = f(p,k)$, and company B produces independent p' and q', then the probability that p'=p is P and the probability that q = q' is also P, and gcd(n,n') factors both moduli with probability 2P(1-P) as was originally the case.

You face a similar failure even if the two companies use the above scheme, but with different functions. Fortunately there is probably not much choice in what function to use. Company

B could of course use $g(p,k) = B + [2^{2k}/p]$ for $B \neq 1$, but this would not be any easier to implement or any more efficient.

But if everybody would all agree to use the same function it would solve the problem.

**Implementation Issues**
For the sake of implementing the multiprecision arithmetic, we point out that $f(p,k) = 1+[2^{2k}/p]$ can be written in the equivalent form

$$f(p,k) = \{2^{2k} + 1 - (2^{2k} \bmod p)\}/p.$$

A. K. Lenstra et. al. suggested the slightly different function

$f(p,k) = \{2^{2k-1} + 1 - (2^{2k-1} \bmod p)\}/p.$ We needed to modify theirs in order to produce a manageable proof of (1) above, so that all of the p's would be greater than all of the q's.

**Does factoring get easier?**
In a 1998 paper [2] with on RSA key generation, A. K. Lenstra discusses a variety of functions of this type and presents credible evidence that the known factoring algorithms will not be able to factor pf(p) any more easily than pq when p and q are independent.

**Can key generation be made O(k)?**
Also in [2] Lenstra discusses functions that generate the keys in time O(k), but it is pretty clear that these function cannot be made bijective, so they won't work for the present purpose. They generate p, compute f(p) and then search forward f(p)+2, f(p)+4,… until reaching g(p) = f(p) + m, the first prime. Any p' for which f(p) < f(p') ≤ f(p) + m will satisfy g(p') = g(p), so there is no hope of making g(p) bijective.

**Another suggestion for making more secure moduli**
Without changing how the random numbers are generated, greater security is achieved by using algorithm 2 below to generate p, rather than algorithm 3 below.

**Algorithm 2**
1. Generate random odd p.
2. Test p for primality.
3. If p fails to be prime goto 1.

**Algorithm 3**
1. Generate random odd p.
2. Test p, p+2, p+4, ... for primality until a prime is found.

Algorithm 3 behaves particularly badly when p is followed by an unusually long stretch of composite numbers. This may account for some of the bad keys found in the experiment of Lenstra et al [1].

**Summary**
1. The use of such a function solves the problem, but only if everyone uses it.
2. Key generation takes time $O(k^2)$ instead of $O(k)$.
3. The resulting moduli seem to be equally hard to factor.